\documentclass[twocolumn,amsmath,amssymb,nofootinbib]{revtex4}

\usepackage{graphicx}
\usepackage[latin1]{inputenc}
\usepackage{dcolumn}
\usepackage{bm}
\usepackage{epsfig}
\usepackage[english]{babel}
\usepackage{blindtext}
\usepackage{lipsum}
\usepackage{xcolor}
\usepackage{hyperref}
\begin{document}

\preprint{APS/123-QED}

\title{Correlations of network trajectories}

\author{Lucas Lacasa$^{1}$}
\email{lucas@ifisc.uib-csic.es}
\author{Jorge P. Rodriguez$^{2}$}
\email{jorgeprodriguezg@gmail.com}
\author{Victor M. Eguiluz$^{1}$}
\email{victor@ifisc.uib-csic.es}
\affiliation{$^{1}$Institute for Cross-Disciplinary Physics and Complex Systems IFISC (CSIC-UIB), Palma de Mallorca (Spain)\\
$^{2}$Instituto Mediterraneo de Estudios Avanzados IMEDEA (CSIC-UIB) 07190 Esporles (Spain)}%

\date{\today}

\begin{abstract}
Temporal networks model how the interaction between elements in a complex system evolve over time. Just like complex systems display collective dynamics, here we interpret temporal networks as trajectories performing a collective motion in graph space, following a latent graph dynamical system. Under this paradigm, we propose a way to measure how the network pulsates and collectively fluctuates over time and space. To this aim, we extend the notion of linear correlations functions to the case of sequences of network snapshots, i.e. a network trajectory. We  construct stochastic and deterministic  graph dynamical systems and show that the emergent collective correlations are well captured by simple measures, and illustrate how these patterns are revealed in empirical networks arising in different domains.
\end{abstract}

\keywords{} \maketitle

Temporal networks \cite{TN_1,TN_2,TN_3} are a mathematically handy way of modelling how different elements in a complex system interact and how such interactions evolve over time. While a substantial research activity has studied how dynamical processes running {\it on} a network --e.g. diffusion, synchronization, epidemics, etc-- are affected when such network backbone is itself dynamically modified \cite{masuda2013temporal,delvenne2015diffusion,Scholtes_natcomm14,lambiotte2013burstiness,van2013non, coevo1, coevo2, coevo3}, 
the programme of studying the network's intrinsic dynamics --the dynamics \textit{of} the network-- has been seldom explored \cite{Grindrod1, Grindrod2,williams,memory}, 
even if such intrinsic dynamics is itself indicative of the interaction dynamics taking place in complex systems. \\ 
Our contention is that, just as complex systems display collective dynamics, temporal networks perform a collective motion in a (high dimensional) phase space --a graph phase space--, rather than being just an aggregation of independently varying links. 
Accordingly, we propose to interpret temporal networks as whole yet not punctual objects performing a trajectory in graph space governed by a latent graph dynamical system \cite{libro_graph_dyn}. Depending on the level of description and the system under study, the dynamical rules by which the graph object evolves over time might be driven by a system's Hamiltonian, by an effective (possibly dissipative) theory, or by stochastic processes. This perspective opens the room to describe how networks collectively pulsate and fluctuate using the solid grounding offered by dynamical systems theory, stochastic processes and time series analysis.\\
\noindent Here we illustrate such a programme by investigating the extension of correlation functions --classically defined to study linear auto- and cross-correlation of signals-- to the case where the object under analysis is a network whose dynamics displays normal modes and develops linear correlations accordingly. Formally, let ${\cal G} = \{G(s)\}_{s=1}^N$ be an ordered sequence of $N$ network snapshots. The index $s$ can be associated with time (hence addressing temporal autocorrelations), space (spatial correlations), or some other property that allows ordering the sequence. Applications of the former case include e.g. contact networks of social \cite{sune} or biological agents which collectively move (active matter) \cite{active} or diffuse \cite{prx} and interact via space proximity which changes over time. The case of spatial correlations relates to understanding how different networks emerging in different spatial regions correlate, an example being to understand how an ecological network varies and adapts to different environments emerging across earth's longitudinal or latitudinal gradients \cite{eco1, eco2, eco3, eco4}.


 For concreteness we consider labelled, unweighted networks with a fixed number of $m$ and $s\equiv t$, i.e.  ${\cal G} = \{A(t)\}_{t=1}^N$, where $A(t)=\{A_{ij}(t)\}_{i,j=1}^m$ is the adjacency matrix of the $t$-th network snapshot. 
Conceptually, the autocorrelation function of an object is the inner product of itself with itself at a later time (the lag), averaged over the dynamics. Accordingly, here we propose to define the network's autocorrelation matrix $\cal{C}(\tau)$ at lag $\tau$ as
\begin{equation}
{\cal C}(\tau) = \frac{1}{N-\tau}\sum_{t=1}^{N-\tau} A(t)A(t+\tau)^\intercal,
\end{equation}
where $A^\intercal$ is the transpose of matrix $A$, and its scalar projection using Frobenius inner product $\langle \cdot,\cdot \rangle_{\text{F}}$, such that
\begin{equation}
c(\tau) = \text{tr}(\cal{C}(\tau)),
\label{projected}
\end{equation}
where $\text{tr}(\cdot)$ denotes the trace operator. It is indeed easy to see that $c(\tau)=\sum_t \langle A(t),A(t+\tau) \rangle_{\text{F}}$, i.e. $c(\tau)$ sums up the components resulting from the Hadamard products (component-wise) of $A(t)$ and $A(t+\tau)$, averaged over all times $t$. The interpretation of $c(\tau)$ is simple: for a temporal network, it computes the autocorrelation (at lag $\tau$) of each edge time series $A_{ij}(t)$ and then sums up all autocorrelations (sum over edges $i,j$). 
The full correlation matrix $\cal{C}(\tau)$ takes into account not only autocorrelations (found in the diagonal of the matrix), but also cross-correlations (found in the off-diagonal terms). More particularly, each off-diagonal term ${\cal C}_{ij}(\tau)$ ($i\ne j$) displays the cross-correlation $\sum_t A_{ik}(t)A_{jk}(t+\tau)$ of pairs of edges forming a path of size 2 between node $i$ and node $j$ that go through an intermediate node $k$, and aggregates this over all intermediate nodes $k$. These off-diagonal terms aggregate the distance-2 temporal dependencies between nodes $i$ and $j$ that emerge as a result of their indirect correlation via the rest of the nodes, and thus provides an effective (mean-field) contribution of the whole network to the net temporal dependence between nodes $i$ and $j$.\\ 
One can further rescale $\cal{C}(\tau)$ by substracting an appropriately computed average of the objects (the adjacency matrices) over the complete trajectory, i.e.
\begin{equation}
\tilde{\cal{C}}(\tau) = \frac{1}{N-\tau}\sum_{t=1}^{N-\tau} [A(t)-\mu]\cdot[A(t+\tau)^\intercal-\mu^\intercal],
\end{equation}
where $\mu=\frac{1}{N}\sum_{t=1}^N A(t)$ is the annealed adjacency matrix of the temporal network. Similarly we also have
$\tilde{c}(\tau)=\text{tr}(\tilde{\cal{C}}(\tau))$ that fulfils 
\begin{equation}
\tilde{c}(\tau) = \sum_t \langle A(t),A(t+\tau) \rangle_{\text{F}}- \langle \mu,\mu \rangle_F
\end{equation}


\section*{RESULTS}
To validate how $\tilde{\cal{C}}(\tau)$ and $\tilde{c}(\tau)$ work, we now define and build temporal network models of benchmark dynamics. 

\begin{figure}[htb]
\centering
\includegraphics[width=0.75\columnwidth]{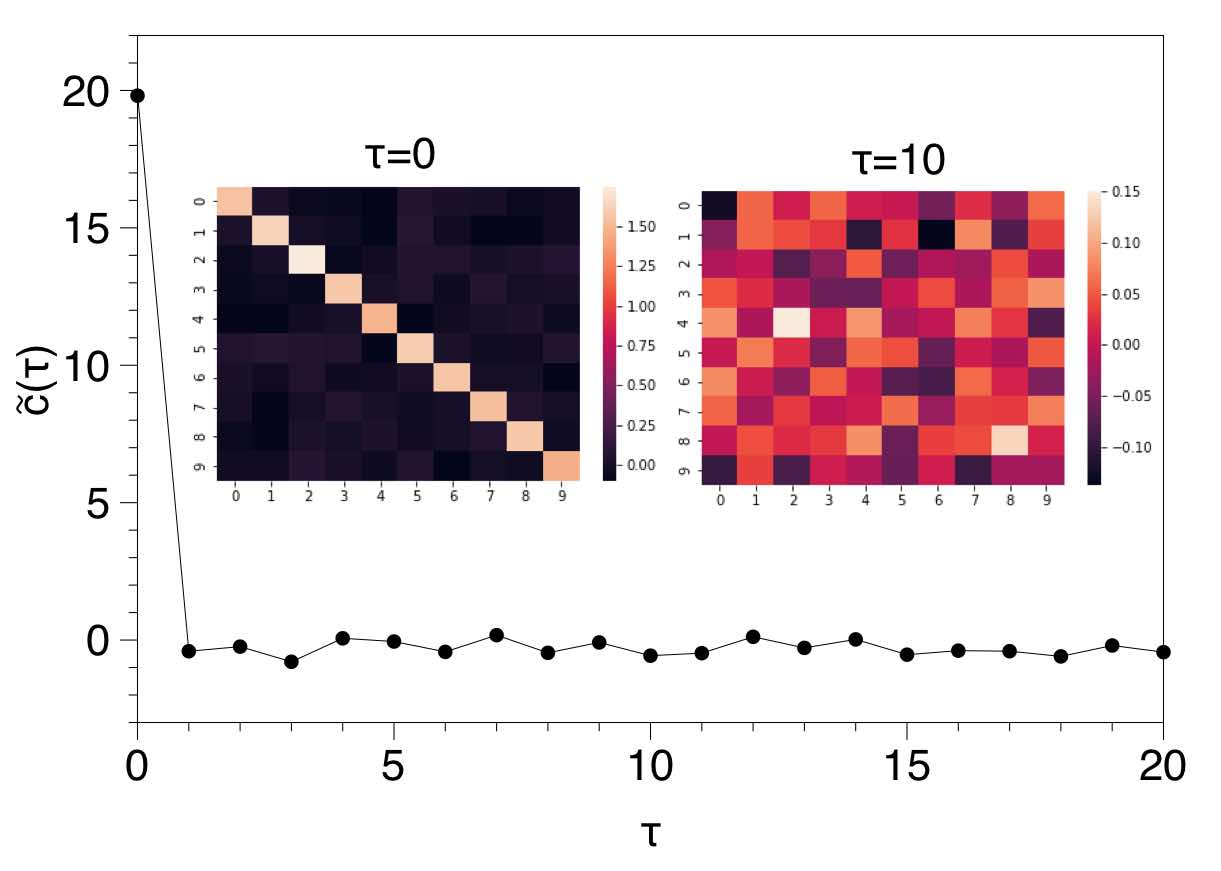}
\caption{\textbf{White temporal networks.} The outer panel shows $\tilde{c}(\tau)$ which evidences the classical Dirac-delta behavior, as expected. The inner panels show $\tilde{\cal{C}}(0)$ and $\tilde{\cal{C}}(10)$, certifying that only autocorrelations emerge, and only at the trivial $\tau=0$.}
\label{fig_white}
\end{figure}

\noindent \textbf{White Networks -- }To build a network version of white noise, we construct an i.i.d. sequence of $N$ Erdos-Renyi graphs ER($p$). In figure \ref{fig_white} we show the result of $\tilde{c}(\tau)$ (outer panel) for $p=0.2$, $m=10$, $N=100$, showing the characteristic  Dirac-delta shape of a white noise's autocorrelation. As inner panels we add heatmaps representing the full correlation matrix at ${\cal C}(\tau=0)$ and ${\cal C}(\tau=10)$, which clearly shows how each of the edges is only autocorrelated at lag $\tau=0$, and no obvious cross-correlation of edges emerges, as expected. One can easily understand this result if we consider the element $A_{ii}(t)A_{ii}(t+\tau)$, which is the building block in Eq.\ref{projected}. By construction links are independent Bernoulli trials, so effectively $\tau$ does not play any role, hence any function depending on $\tau$ which is based on  $A_{ii}(t)A_{ii}(t+\tau)$ will be necessarily flat for $\tau>0$.\\

\begin{figure*}[htb]
\centering
\includegraphics[width=0.85\columnwidth]{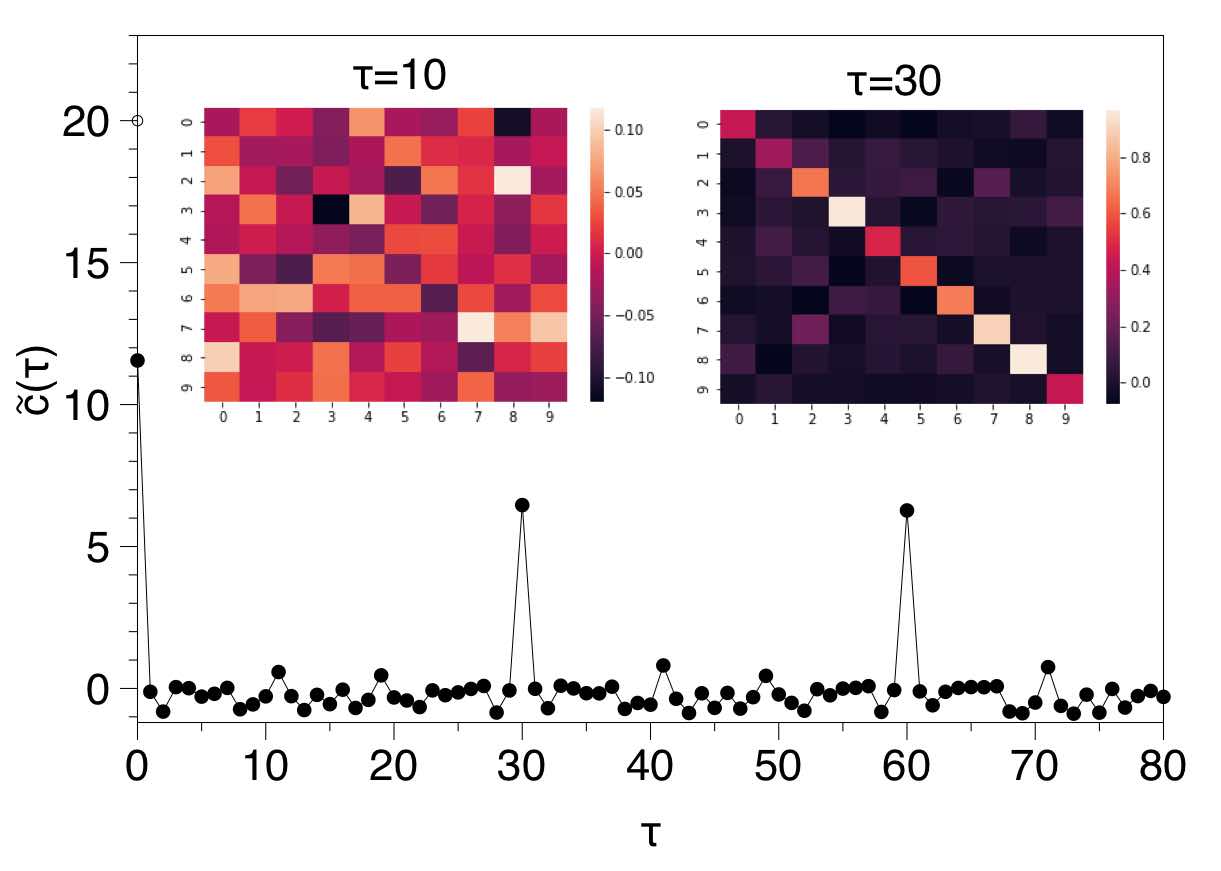}
\includegraphics[width=0.85\columnwidth]{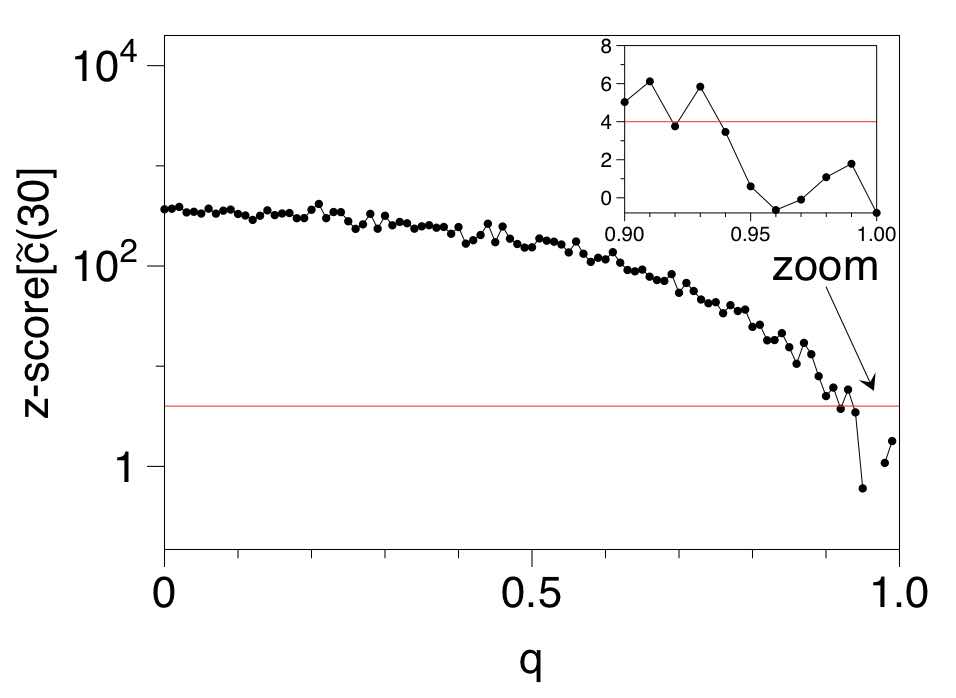}
\caption{\textbf{Noisy periodic networks.} The outer panel shows $\tilde{c}(\tau)$ which evidences the peaks at multiple harmonics of the period, as expected. (Left panel) The inner panels show $\tilde{\cal{C}}(10)$ and $\tilde{\cal{C}}(30)$. (Right panel) Period detectability, for which we compute the z-score statistic associated to $\tilde{c}(30)$ (see the text). As a guide, we highlight the detectability threshold z-score$=4$, for which the deviations between $\tilde{c}(30)$ and the average inter-period signal $\tilde{c}(1...29)$ is 4 times larger than expected by chance, and conclude that the hidden periodicity can be captured even for very high levels of noise.}
\label{fig_periodic}
\end{figure*}

\noindent \textbf{Noisy periodic networks--} On a second step, we build periodic networks of period $T$, 
by first constructing an i.i.d. sequence of $T$ Erdos-Renyi graphs ER($p$) and then concatenating several of these sequences one after the other to build the temporal network with $N$ snapshots. To make the quantification of periodicity more challenging, we pollute the (pure) periodic temporal network pattern with a certain amount of noise: each edge is independently affected by noise with probability $q$., and those edges affected by noise are set to 1 with probability $p$ and to 0 with $1-p$ (by construction, the periodic pattern is completely washed out for $q\to1$). In figure \ref{fig_periodic} illustrates the result for $(m,T,N,p,q)=(10,20,120,0.1,0.4)$. The temporal network shows a clear periodic pattern at $T$ as its autocorrelation function peaks at $\tau=T$ and successive harmonics, and there is no trace of cross-correlation or autocorrelations at $\tau\neq T$, as expected. Inner panels depict 
${\cal C}(\tau=T)$ and ${\cal C}(\tau \neq T)$, indicating that correlations are only found when $\tau$ is a multiple of $T$. To analytically justify this behavior, we can again resort to analysing $A_{ii}(t)A_{ii}(t+\tau)$. This binary quantity is 1 with probability $p^2$ for all $\tau \ne T$, and is 1 with probability $p$ when $\tau=0,T,2T,...$ Using expected values, since $p>p^2$, $\tilde{c}(\tau)$ will pulsate at $\tau=T$ and its harmonics, and will have a much smaller value for other values of $\tau$.\\
Now, we expect that as the level of noise increases, this relation does not necessarily hold for all nodes, eventually breaking up the pulsation. In order to quantify the periodic detectability, we compute the $z$-score of $\tilde{c}(T)$, defined\footnote{one can easily extract a $p$-value from this $z$-score, the null hypothesis being that, for a given noise level $q$, the observed value $\tilde{c}(T)$ is not sufficiently high to determine in a statistically significant way that there is indeed a period at $\tau=T$} as $\frac{\tilde{c}(T) - \langle\{\tilde{c}(1), \tilde{c}(2),\dots, \tilde{c}(T-1)\}\rangle}{\sigma(\{\tilde{c}(1), \tilde{c}(2),\dots, \tilde{c}(T-1)\})}$. In the right panel of Fig.\ref{fig_periodic} we plot such $z$-score as a function of $q$, for $T=30$. Assuming a detectability threshold of 4 (rejecting the null hypothesis with very large confidence), one can assert that the noisy network has a periodic backbone up to very high noise levels $q\approx0.9$ (this property surely depends on other parameters such as the number of nodes $m$, the wiring probability $p$, etc).\\

\begin{figure}[htb]
\centering
\includegraphics[width=0.65\columnwidth]{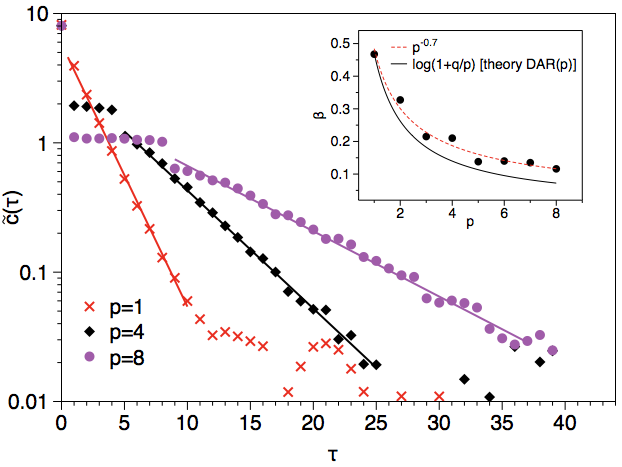}
\caption{\textbf{DARN(p).} Semi-log plot of $\tilde{c}(\tau)$ for a DARN($p$) of $m=10$ nodes and $N=10000$ snapshots, with parameters $q=0.6$, $y=0.1$. The network displays constant correlation for $\tau \le p$, and decays exponentially fast thereafter as $\exp(-\beta \tau)$. The inset panel describes the relation between the decaying exponent $\beta$ and the order of memory $p$, and a the best fit gives $\beta \sim p^{-0.7}$.}
\label{fig_DARNP}
\end{figure}

\begin{figure}[htb]
\centering
\includegraphics[width=0.85\columnwidth]{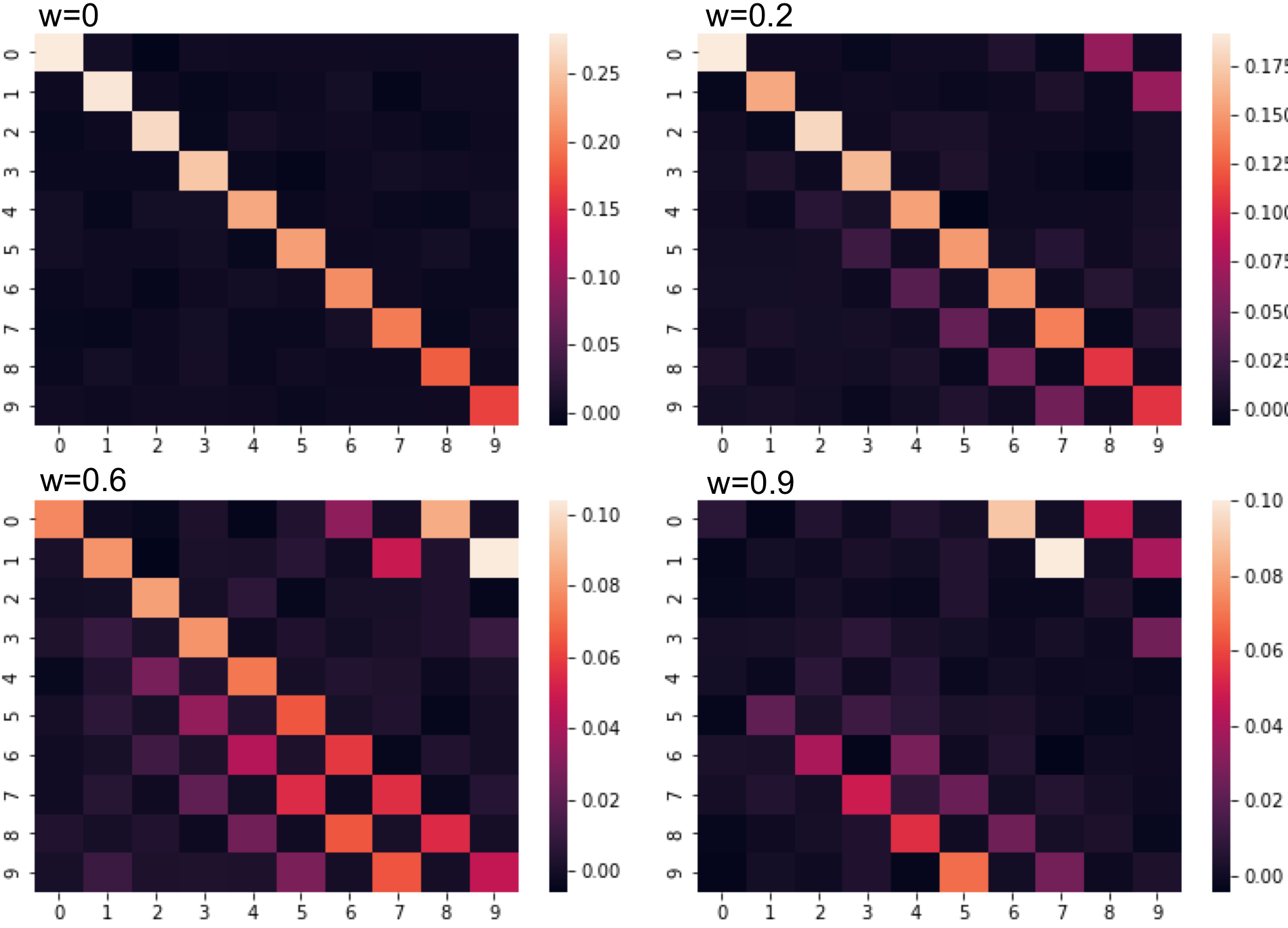}
\caption{Full heatmap plots of $\tilde{{\cal C}}(\tau)$ for a modified DARN($1$) model $m=10$ nodes and $N=10000$ snapshots, with parameters $q=0.6$, $y=0.1$, where with probability $w$, when the link update $A_{ij}$ is from its past, we instead update it with the past of $A_{ii'}$, $j'=j+2 \mod(m)$. When $w=0$, the model is just a standard DARN($1$). As $w$ increases, the links increasingly develop cross-correlations at the expense of auto-correlations.}
\label{fig_DARNP_bis}
\end{figure}

\noindent \textbf{Memory --} As a third step, we now generate synthetic models of temporal networks with prescribed memory. We first consider so-called a discrete autoregressive network models or DARN($p$) \cite{williams, memory}, which is the network version of a discrete autoregressive process of finite order $p$ \cite{thesis_oli}. In this model, each link evolves independently, and at each time step either makes a copy of its state from its past (taking the copy from a random position of its past $p$ states), or updates randomly. Formally, the dynamics of a single link follows $\ell_t = Q \ell_{t-Z} + (1-Q) Y$, where $Q$ is either 0 or 1 (Bernoulli trial), $Z$ is a random variable that draws values from $\{1,2,\cdots,p\}$, and $Y$ is again a binary random variable that results from another Bernoulli trial. This model generates binary values for each link $\ell$ and one can prove that overall the process is non-Markovian, with order $p$. In Fig.\ref{fig_DARNP} we plot the values of  $\tilde{c}(\tau)$ for different memory orders $p$. We observe that correlation is constant for $\tau \le p$ and seems to have an exponential decay thereafter. The rate of decay itself decreases when the memory order $p$ increases, as shown in the inset panel of the same Figure (the solid line corresponds to an analytical result on DAR($p$) processes which involves a local approximation and is only valid when $\beta$ is fitted in $p+1<\tau<2p$ \cite{thesis_oli}). Overall, results suggest that $\tilde{c}(\tau)$ adequately captures the linear temporal correlations of the network.\\
To complete this example, we now relax the assumption that each edge samples its future state from its own past and allow, with a certain probability $w$, that such sampling is performed from the past of a different link. This induces non-negligible cross-correlations and, as a result, the network pulsation is more complex. In this scenario, $\tilde{c}(\tau)$ does not capture all the macroscopic temporal correlations, and one needs to consider the full correlation matrix $\tilde{\cal{C}}(\tau)$. We illustrate this effect in Fig.\ref{fig_DARNP_bis}, where we can appreciate that, as the probability $w$ increases, off-diagonal terms emerge, and eventually take over the diagonal ones for large values of $w$.\\

\begin{figure}[htb]
\centering
\includegraphics[width=0.8\columnwidth]{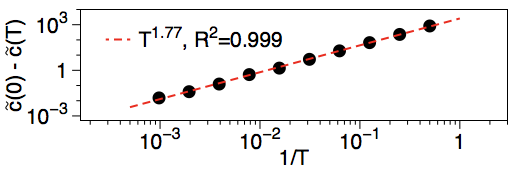}
\includegraphics[width=0.85\columnwidth]{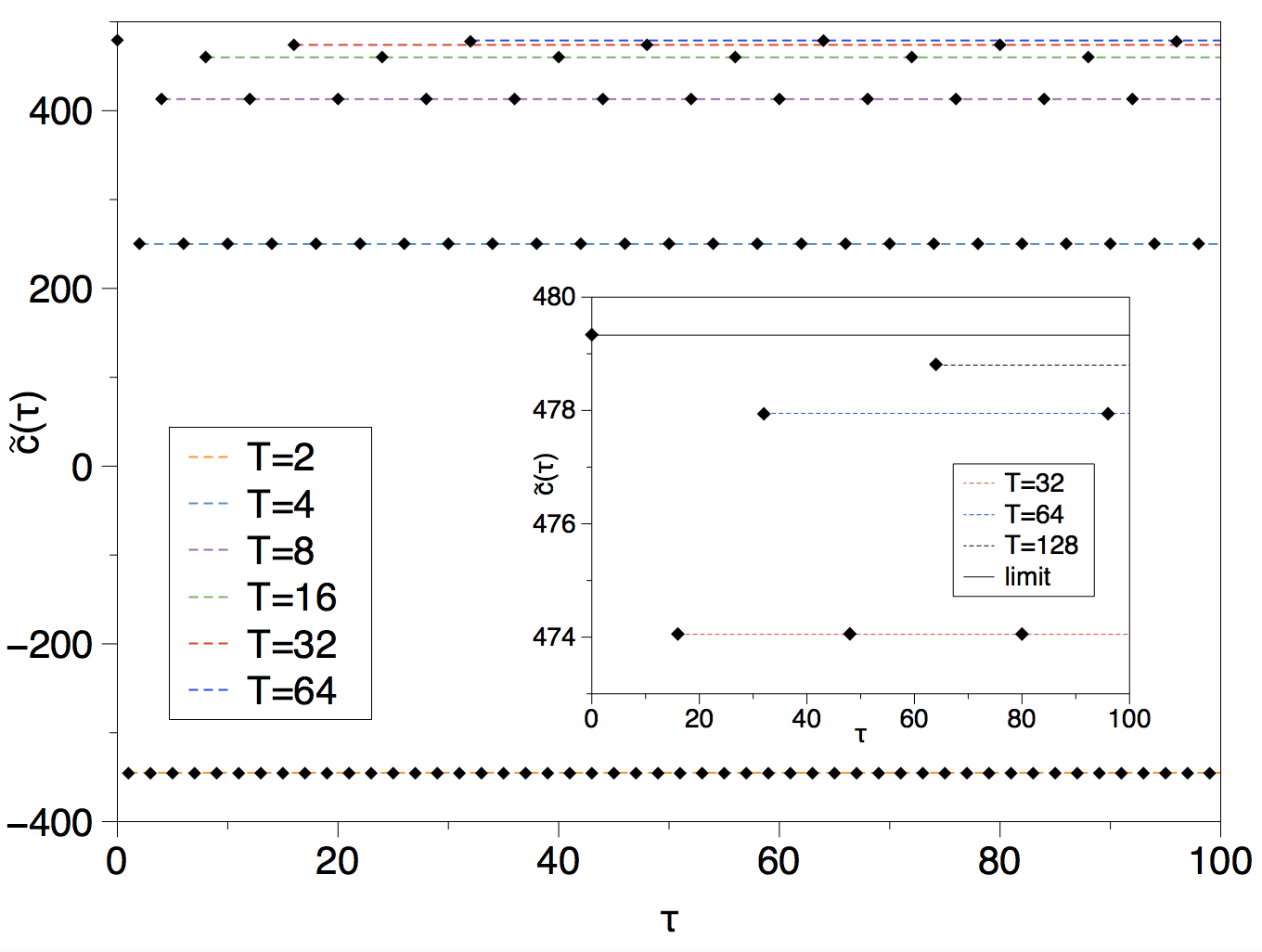}
\caption{\textbf{Edge of chaos.} $\tilde{c}(\tau)$ for a logistic temporal network  of $m=100$ nodes and $N=10000$ snapshots poised at the edge of chaos ($r\approx r_{\infty}$). The network trajectories display a fractal structure with infinitely-many periods (hierarchically organised as powers of 2), captured as periodic peaks of different heights by the correlation function. The top panel provides a scaling relation between such heights and the period they relate to, with a non-trivial exponent.}
\label{fig_acc}
\end{figure}

\noindent \textbf{Edge of chaos --} We now proceed to construct deterministic temporal networks with complex dynamics, including chaos and fractality. To construct `chaotic networks' we initially generate a `dictionary' of networks ${\cal D}=(G_1,G_2,...,G_L)$ such that $||G_p-G_q||=|p-q|$, for some chosen network norm $||\cdot||$. The dictionary is generated sequentially with $G_1 \sim \text{ER}(p)$ and constructing $G_{\ell+1}$ by rewiring a link which (i) had not been rewired before, (ii) into a place that did not have a link before. It is easy to see that ${\cal D}$ is provides a partition of any one-dimensional interval (we take $[0,1]$ without loss of generality) as $[0,1]=\cup_{\ell=0}^{L-1} (\ell/L, (\ell+1)/L]$, so that one can generate network trajectories out of unit interval dynamics by matching $(\ell/L, (\ell+1)/L] \to G_{\ell+1}$.
For illustration, we choose
the logistic map $x_{t+1}=rx_t(1-x_t), \ 0<r\le4, \ x\in[0,1]$. This map generates a period-doubling cascade of signals with period $T=2^k$ as $r$ increases, with a period diverging at a finite $r_{\infty}\approx 3.5699456$. For  $r>r_{\infty}$ the map produces chaotic trajectories  intertwined with other routes to chaos.\\ 
We analysed $\tilde{c}(\tau)$ for two interesting cases: $r=4$ (fully-developed chaos) and $r=r_{\infty}$ (edge of chaos), for  $L=1000$ and networks with $m=100$ nodes and $p=0.4$ The $r=4$ case is indistinguishable from the case of white networks (Fig.\ref{fig_white}), as expected given that fully-developed chaos lacks linear correlations. The $r_\infty$ case is reported in Fig.\ref{fig_acc}, finding a rich, self-similar correlation structure with an intertwined hierarchy of periodically-separated peaks, reminiscent of the infinitely-many modes with period $T=2^k$ of the dynamics at the edge of chaos. The height of these peaks increases as a function of the specific mode $k$ in such a way that, when looking at how the correlation peaks approach $\tilde{c}(\tau=0)$ (in units of the correlation function), we unveil a scaling $\tilde{c}(0)-\tilde{c}(T)\sim T^{-\alpha}$, with a nontrivial exponent $\alpha \approx 1.77$, which reminiscent of --although not obviously related to-- Feigenbaum constants.\\


\begin{figure*}[htb]
\centering
\includegraphics[width=1.9\columnwidth]{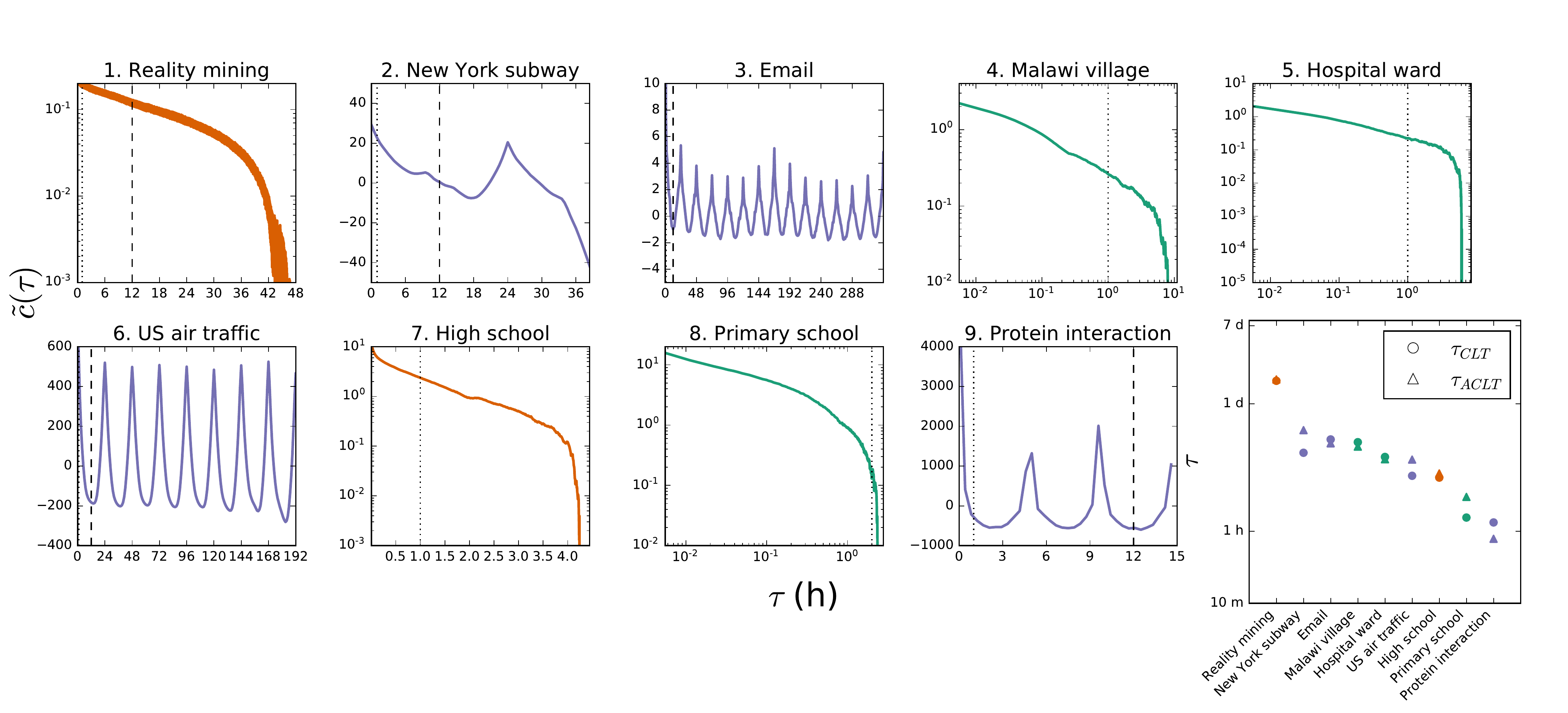}
\caption{\textbf{Empirical networks.} Plots of $\tilde{c}(\tau)$ as a function of $\tau$ (hours) for different empirical temporal networks (see the text and appendix for details). Some of the panels are in semi-log scale, some of them in log-log scale, some of them in linear scales. The dotted and dashed dotted vertical lines correspond to $\tau=1$ and $12$ hours respectively. Across these systems we find a wide range of emerging stylized correlation patterns that match the prototypical structures found in the synthetic models, from pure periodicity --highlighting temporally pulsating networks-- (blue) to both short-range (orange) and long-range correlation structures (green). The bottom-right panel depicts the correlation life-time ($\tau_{CLT}$), and  activity-preserved correlation life-time ($\tau_{ACLT}$), defined as the first time after which the curves hit $\tilde{c}=0$ and the first time after which the curve crosses a properly shuffled null model, respectively (these quantities are well defined only for decaying correlation curves, not so for periodic ones). For a deeper analysis of internal correlation structure, see the full correlation matrices (SI). We find that social interaction-based networks typically display linear correlations at the network size that decay with different speeds (exponentially or with a power law decay) but don't show evidence of harmonicity, suggesting that the underlying complex social system evidences different degrees of memory. This contrasts with online interaction (emails), that follow a markedly regular pattern with a collective periodicity of $T\approx 1$ day, with a second periodic mode showing up at $T \approx 1 $ week. Such 
periodic structure --possibly reminiscent of an underlying scheduling-- is also found in origin-destination flows found in both subway and air transport (daily periodicity). More surprising is to observe that periodic modes also emerge in a biological system such as a protein interaction network: here we speculate that the observed periodicity is related to the typical length of a full metabolic cycle of $\approx 5$ hours \cite{pin, metabolic}.}
\label{fig_empirical}
\end{figure*}

\noindent \textbf{Empirical  networks -- } To round off, we now apply our methodology to a range of different empirical temporal networks that characterize evolving interaction patterns in different systems, including online  (email networks \cite{email}) and offline social interaction in different settings (proximity networks in a university \cite{remine}, a hospital ward \cite{hospital}, primary \cite{primaryschool} and high school \cite{highschool}, interactions in a village \cite{malawi}), transportation networks (NY subway, US air traffic \cite{nets}), and biological systems (protein interactions \cite{pin}), see appendix for details. Across these systems we find a wide range of emerging stylized correlation patterns that match the prototypical structures found in the synthetic models, from pure periodicity --which highlights temporally pulsating networks-- to both short-range and long-range correlation structures. Values of $\tilde{c}(\tau)$ for all systems are plotted in Fig.\ref{fig_empirical}. We find that social interaction-based networks typically display linear correlations at the network size that decay with different speeds but don't show evidence of harmonicity, suggesting that the underlying complex social system evidences different degrees of memory. This contrasts with online interaction (emails), that follow a markedly regular pattern with a collective periodicity of $T\approx 1$ day, with a second periodic mode showing up at $T \approx 1 $ week. Such 
periodic structure --possibly reminiscent of an underlying scheduling-- is also found in origin-destination flows found in both subway and air transport (daily periodicity). More surprising is to observe that periodic modes also emerge in a biological system such as a protein interaction network: here we speculate that the observed periodicity is related to the typical length of a full metabolic cycle.\\
Overall, these results point to the fact that empirical temporal networks indeed describe collective fluctuations which can be captured and interpreted using the network extension of a simple linear correlation formalism. The correlation matrices provide a more microscopic insight into internal correlations across links and in SI Fig S1 we plot these for different lags $\tau$ of interest detected by $\tilde{c}(\tau)$.



\section*{DISCUSSION}
We have presented a parsimonious way to capture the correlation structure in a sequence of networks by interpreting this sequence as a trajectory of a latent graph dynamical system. For a (complementary and) microscopic analysis of memory, we in turn refer the readers to \cite{memory}. As previously stated, the measure proposed here can equally explore temporal correlations (when the ordering index $s$ represents time) or spatial correlations (when $s$ represents space). In the latter case, $s$ represents either a one dimensional direction (e.g. latitudinal or longitudinal variations) or a generalised (e.g. a radial) one. Applications pervade physical systems e.g. condensed matter (fluctuating spin lattices), biophysics (active matter contact networks) or quantum physics (quantum networks), and includes areas beyond physics such as social mobility, technological transportation networks, or biological systems. This measure can be also trivially extended to assess cross-correlations between two sequences of networks.\\ 
Further research should also consider the case where (i) each network snapshot has a vertex set with potentially different cardinality (i.e. the vertex set size also fluctuates over time or space), or where (ii) nodes don't have an explicit label. Problem (i) can be tentatively addressed by sampling the maximal subgraph that contains a unique vertex set. Problem (ii) is more computationally challenging and relates to the problem of canonically embedding an unlabelled network into a labelled one, a possible solution (computationally affordable for medium-size networks) relates to label nodes via Haussdorff-Gromov embeddings.\\
In this paper we deliberately choose an intuitive property --linear correlations-- to convey our idea of interpreting temporal networks as trajectories. The broader research programme extends above and beyond correlations: we envisage other dynamical properties (e.g. dynamical stability)  to be similarly extended to analyse network trajectories.

\section*{Appendix: empirical networks}
\begin{itemize}
\item The Malawi village \cite{malawi}, hospital ward \cite{hospital}, high school \cite{highschool}, and primary school \cite{primaryschool} temporal networks were publicly available from the Sociopatterns collaboration. These networks represent the face-to-face contacts between human individuals that are tracked through wearable sensors that report their mutual proximity detections, with a resolution of 20 s. In particular, the Malawi village tracked 86 individuals living in the same village, the hospital ward 46 health care workers and 29 patients, the high school 329 students, and the primary school 232 students and 10 teachers.
\item The reality mining network \cite{remine} tracked the proximity contacts between 75 students/faculty in the MIT Media Laboratory and 25 incoming students at the MIT Sloan business school using the Bluetooth technology to detect proximity between the phones carried by the tracked individuals. Although the contacts have the resolution of 1 second, we grouped them to an unweighted undirected network with a resolution of 20 s to smooth the noise oscillations in the lower time scales.
\item 
The New York subway network \cite{nets} represents the timetabled journeys for two days, connecting origin and destination stations of each of the 660 K possible trips between 417 different stations, with the resolution of 2 min. We created an undirected unweighted network connecting two stations if at least one trip was observed between them with a resolution of 30 min. 
\item 
The email network \cite{email} included the information about the senders and receivers of the email leaks of the Democratic National Committee in 2016, with a resolution of 1 s. It included two periods, one with lower reported activity and, after 466 days, a final period with a higher activity, upon which we performed our analysis, creating an unweighted undirected temporal network with a resolution of 1 h that connects two individuals if at least one email was exchanged between them within our resolution period, with a total of 1629 individuals. 

\item 
The US air traffic network \cite{nets} reported the flights between 299 US airports for 10 days with a resolution of 30 min. We created an unweighted undirected temporal network connecting two airports if there was a flight between them, with a resolution of 1 h.
\item 
The protein network \cite{pin}  included the protein-protein interactions on yeast cells. There were 12 available networks from different experiments, and we selected that with the highest number of nodes (DPPIN-Babu, 5003 nodes), reporting 111 K interactions with a resolution of 25 minutes, including the information through 3 consecutive metabolic cycles.

\end{itemize}


\begin{acknowledgements}
The authors thank Oliver Williams, Konstantin Klemm and Juan Fernández-Gracia for insightful discussions. LL acknowledges funding from project DYNDEEP (EUR2021-122007),  and LL and VME acknowledge funding from project MISLAND (PID2020-114324GB-C22), both projects funded by Spanish Ministry of Science and Innovation. \\
\noindent \textbf{Code availability. } Upon publication, Python implementations of all algorithms will be available at  \url{https://github.com/lucaslacasa/}.
\end{acknowledgements}

\bibliography{apssamp}

\end{document}